\documentclass[twocolumn,floats,aps]{revtex4}
\usepackage{times,amsmath,psfrag,latexsym,pstricks,graphics}

\begin{document}
\title{Anisotropic Dependence of Superconductivity on Uniaxial Pressure in
CeIrIn$_5$}
\author{O. M. Dix, A. G. Swartz, R. J. Zieve}
\affiliation{Department of Physics, UC Davis} 
\author{J. Cooley}
\affiliation{Material Science and Technology Division, Los Alamos National Laboratory}
\author{T. R. Sayles$^*$ and M. B. Maple}
\affiliation{Department of Physics, UC San Diego}

\begin{abstract}
We measure the effect of uniaxial pressure on the superconducting
transition temperature $T_c$ in CeIrIn$_5$.  We find a linear change in
$T_c$ with both $a$-axis and $c$-axis pressure, with slopes of 56 mK/kbar
and -66 mK/kbar, respectively.  By comparing results from doping studies
and different types of pressure measurements, we separate the influences
of hybridization and dimensionality on $T_c$.  We find the true geometric
influence, for constant hybridization, is $\partial T_c/\partial(c/a)=44$
K.

\end{abstract}
\maketitle

The low-temperature phases of heavy fermion materials have attracted much
attention in recent years.  They exhibit a range of correlated phases, including
several types of magnetism.  Superconducting regimes emerge near
zero-temperature magnetic phase transitions.  Non-Fermi liquid behavior also
appears near these quantum critical points, and can persist to significant
temperatures.  Tuning through alloying, pressure, or applied field allows
exploration of the exact balance among the phases.

One of the most-studied heavy fermions recently has been CeMIn$_5$
(M = Ir, Rh, Co).  The proximity of these Ce-based 115 compounds to an
antiferromagnetic quantum critical point leads to a rich phase diagram
\cite{pham2006}.  CeRhIn$_5$ at ambient pressure has a superconducting
transition near 0.1 K, deep within an antiferromagnetic phase \cite{ambientRh}.
Hydrostatic pressure destroys the magnetism and raises $T_c$ to a maximum  of
2.1 K at 16 kbar \cite{hegger, thompson2001}. On the other hand, CeIrIn$_5$ and
CeCoIn$_5$ superconduct at ambient pressure \cite{petrovicir, petrovicco}, with
several similarities to the high-temperature cuprate superconductors.  The 115
materials have a tetragonal, HoCoGa$_5$ crystal structure which can be viewed
as alternating layers of CeIn$_3$ and MIn$_2$ stacked along the (0 0 1)
direction \cite{petrovicir, hegger, petrovicco}, reminiscent of the
copper-oxygen planes in the cuprates. Power-law temperature dependences in heat
capacity, thermal conductivity and spin-lattice relaxation rate in the
superconducting state suggest $d$-wave symmetry of the superconducting order
parameter \cite{thompson2003}. Furthermore, emergence of superconductivity near
$T_N$ and coexistence of homogeneous antiferromagnetism and superconductivity
provide evidence that magnetic fluctuations mediate Cooper-pairing in these 
systems \cite{pagliuso, mito, kawasaki}.

One major influence on the superconducting transition temperature is
the hybridization between the Ce $f$-electrons and In $p$-electrons
\cite{thompsoniso,rkumar702004,fujimori}, which controls the spin-fluctuation
temperature, $T_{sf}$.  $T_{sf}$ is proportional to $T_c$ and roughly inversely
proportional to $\gamma$, the Sommerfeld coefficient of the normal-phase
specific heat.  Smaller $\gamma$'s in the isostructural PuMGa$_5$ coumpounds (M
= Rh, Co) partially account for their much higher superconducting transition
temperatures.  However, even within the CeMIn$_5$ series, $\gamma$ alone cannot
describe all the variation in $T_c$; for example, while $\gamma$ changes by
just over a factor of 2 between M = Ir and M = Co, $T_c$ changes by nearly a
factor of 6.  Thus hybridization cannot be the sole influence on $T_c$.

Mean-field theoretical models of magnetically mediated superconductivity
indicate a strong dependence of $T_c$ on dimensionality \cite{mathur,
monthoux2001, monthoux2002}.  Measurements on CeM$_{1-x}$M$^{\prime}_x$In$_5$
show a linear relationship between $T_c$ and the ratio of the tetragonal
lattice constants $c/a$ \cite{pagliuso}.  Interestingly, for the Pu-based 115
materials $T_c$ is also linear in $c/a$, with the same relative change in $T_c$
with dimensionality, $\frac{1}{T_c}\frac{dT_c}{d(c/a)}$ \cite{thompsoniso}. 
The agreement in slopes makes sense if the difference in hybridization sets the
overall temperature scale for each family but dimensionality governs the
behavior within each family.  Another way to control dimensionality and
hybridization is by applying pressure.  Under hydrostatic pressure, $c/a$ is
not even monatonic for CeRhIn$_5$ and CeCoIn$_5$, and $T_c$ is not linear in
$c/a$.   However, all the CeMIn$_5$ compounds have similar hybridization at the
pressure $P_{max}$ which maximizes the $T_c$.  Considering $T_c$ and $c/a$ at
$P_{max}$, where the effects of hybridization differences are reduced, does
give  a linear relationship \cite{rkumar692004, rkumar702004}. Kumar {\em et
al.} argue that hydrostatic pressure mainly alters the hybridization.  The
similar hybridizations at $P_{max}$ suggest that hybridization determines the
pressure $P_{max}$ while dimensionality governs the value of $T_c$ at $P_{max}$
\cite{rkumar702004}.

Uniaxial pressure is a natural technique for further exploring the effects of
dimensionality, since the $c/a$ ratio of an individual sample can be
increased or decreased depending on the pressure axis.  Uniaxial pressure
leads to fairly small changes in hybridization, since lattice constants
decrease along the direction of applied force but increase in the
perpendicular directions.  On the other hand, the effects on $c/a$ are
much larger than for a similar hydrostatic pressure.  To date,
the effects of uniaxial pressure on the $115$ materials have not been
measured directly, although thermal expansion measurements combined with
Ehrenfest relations predict the change in $T_c$ with uniaxial pressure in
the zero pressure limit \cite{oeschler}.  Here we explore the dependence
of $T_c$ on pressure and hence on $c/a$, and combine our results with those
on hydrostatic pressure and alloying to extract the dependence of $T_c$ on
dimensionality.

Samples were grown in alumina crucibles from a molten metal flux containing
stoichiometric amounts of Ce and Ir, and excess indium.  We confirmed through
transmission Laue X-ray diffraction that the lattice constants of our samples
matched those reported elsewhere \cite{yoshinori}.  

\begin{figure}[t!hbp]
\begin{center}
\psfrag{RuO$_2$}{\scalebox{2.5}{RuO$_2$}}
\scalebox{.34}{\includegraphics{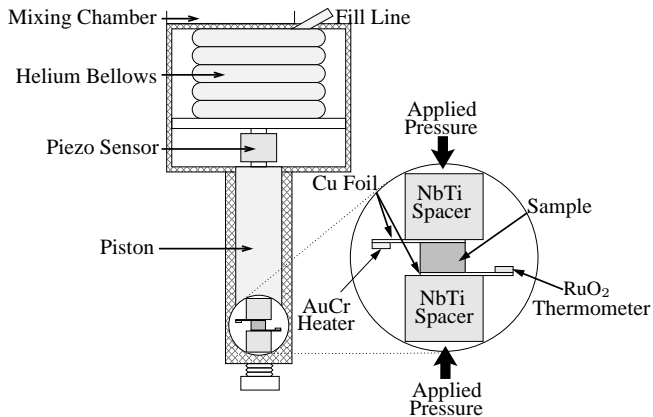}}
\caption{Helium bellows setup for measuring heat capacity under uniaxial
pressure.}
\label{presscell}
\end{center}
\end{figure}

Our uniaxial pressure apparatus, shown in Figure \ref{presscell}, is a
helium-activated bellows mounted on a dilution refrigerator
\cite{ube13}, and permits changes in pressure without thermally cycling the
sample.  We monitor the pressure through a piezoelectric crystal in the
pressure column.  The maximum achievable pressure depends on the size of the
sample, but is typically about 10 kbar.
We measure the superconducting transition with adiabatic heat capacity.
Superconducting NbTi spacers between the sample and the pressure cell
serve as the thermal link.

As can be seen from Figure \ref{presscell}, the sample is unconstrained in
directions perpendicular to applied pressure.  To apply
pressure along a specific axis, we first orient our sample using Laue
X-ray diffraction.  We then polish the sample according to the desired
orientation.  We preserve as much bulk as possible during the polishing, since
the observed time constant for thermal decay is proportional to the sample's
mass.  Our largest samples for pressure
along the $c$ and $a$ axes were approximately 30 mg and 80 mg,
respectively. 

\begin{figure}[b!h]
\begin{center}
\scalebox{.4}{\includegraphics{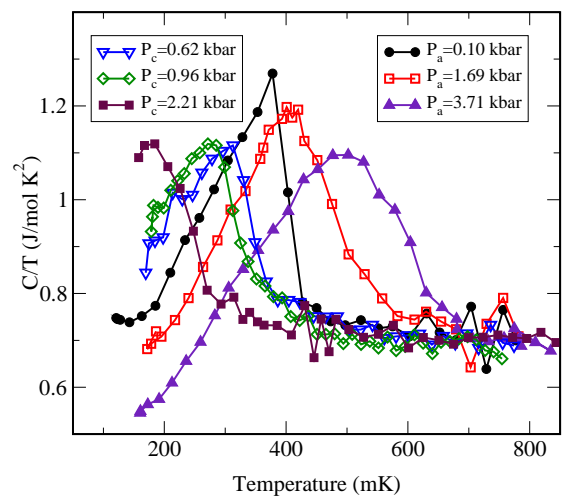}}
\caption{Heat capacity at various pressures.
The data have been scaled by the normal phase heat capacity; see text.}  
\label{covrtvst}
\end{center}
\end{figure}

Figure \ref{covrtvst} shows representative heat capacity data.  Pressure
applied along the $c$ axis shifts the transition to lower temperature, while
$a$-axis pressure has the opposite effect.  The data are scaled, with a single
scaling constant used for each sample, regardless of pressure. The average
$C/T$ over all pressures at 800 mK is set to 700 mJ/mol K$^2$.

For $a$-axis pressure, we fit a function $\gamma_s+a_1T^n+a_2T^{-3}$ to
the heat capacity in the superconducting phase \cite{petrovicir}.  The
form fits well at all pressures, with the exponent $n$ always close to 1,
as expected for line nodes in the energy gap.  For $c$-axis pressure, the
reduced $T_c$ leaves too small a temperature range for reliable fits.  If
we fix $n=1$ and refit the data, we find a slight decrease in both
$\gamma_s$ and $a_1$ with pressure, as opposed to the sharp change in
$\gamma_s$ observed with uniaxial \cite{ube13} and hydrostatic 
\cite{Caspary, Fisher02} in other heavy fermion superconductors.

\begin{figure}[b!h]
\scalebox{.4}{\includegraphics{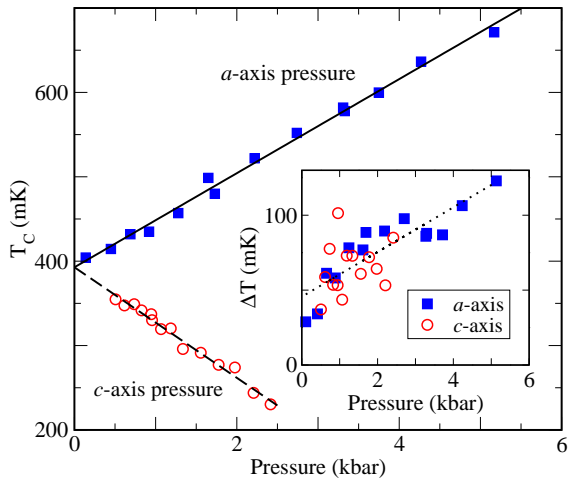}}
\caption{Transition temperature as a function of uniaxial pressure.
$T_c$ was determined using the temperature corresponding to the average
of the $C/T$ values at the peak and onset of the transition.  Lines are
linear fits, with slope 56 mK/kbar and -66 mK/kbar.  Inset: 20\%-80\%
transition width vs. pressure.  The dotted line is a guide to the eye.}
\label{tcvsp}
\end{figure}

We define $T_c$ as the temperature where $C/T$ equals the average of its
maximum value in the superconducting region and its value in the normal
phase at the onset of the transition. We also model the transition as an
abrupt discontinuity with the restriction that the normal-phase entropy
is the same as for the actual data.  The two methods give consistent
values of $T_c$.  However, under $c$-axis pressure $T_c$ decreases until
eventually there is not enough data in the superconducting region to
extrapolate the value of $C/T$ in an equal-entropy calculation.  For
consistency between $c$ and $a$-axis pressure, all values we report here
use the average-$C/T$ method for $T_c$ in both pressure directions.  The
transition temperatures for different pressures, obtained using this
average $C/T$ method, are shown in Figure \ref{tcvsp}.  In our pressure
range, the change in $T_c$ is linear.  It equals 56 mK/kbar for $a$-axis
pressure, -66 mK/kbar for $c$-axis pressure.  These values agree fairly
well with those derived for the zero-pressure limit from thermal
expansion data: 54 mK/kbar and -89 mK/kbar, respectively \cite{oeschler}.
For the tetragonal crystal structure, we can also compare our results to
the effect of hydrostatic pressure through $\partial T_c/\partial
p_V=2\partial T_c/\partial p_a + \partial T_c/\partial p_c$.  Our data
yield a value of $\partial T_c/\partial p_V=46$ mK/kbar, somewhat larger
than the 25 mK/kbar obtained through direct hydrostatic pressure
measurements \cite{hydrostat}.

We also comment on the normal state heat capacity. We find no significant
change with $a$-axis pressure but an increase $0.03\pm0.01$ J/kbar mol K$^2$
with $c$-axis pressure.  Previous hydrostatic pressure measurements
\cite{hydrostat} found a decrease of -0.02 J/kbar mol K$^2$.
Although the sign differences here and in $\delta T_c/\delta P$ may be related,
the magnitudes of the heat capacity changes
do not correspond to the overall Ce-In hybridization for the different
types of pressure.  Instead they suggest that hybridization in certain
directions affects the heat capacity more than in others. 
The superconducting jump averages $\Delta C/\gamma T_c=0.73\pm0.05$
J/mol K$^2$ and has no apparent trend with pressure, consistent with
previous experiments \cite{petrovicir,yoshinori,hydrostat}.

We take as the width of the transition the span between the temperatures
corresponding to 20\% and 80\% of the $C/T$ range during the transition, as
shown in the inset of Figure \ref{tcvsp}.  Under pressure the transition
becomes broader and more rounded.  Although an inhomogenous pressure
distribution across the face of the sample or a variation in the
cross-sectional area over the height of the sample would broaden the
transition, reasonable values for these effects would produce less than 15\%
of the observed broadening.

We apply a small pressure during cooldown to keep the sample in place.
This initial pressure differs for each
cooldown but is typically on the order of 0.3 kbar.  To determine the initial
pressure, we extrapolate $T_c$ versus $p$ to where the transition temperature
reaches that measured on samples outside of the pressure cell.  The
pressure labels for the curves in Figure \ref{covrtvst} are adjusted for
the initial pressure, so that the values listed are the correct pressures.

\begin{figure}[t]
\begin{center}
\psfrag{c/a}{\scalebox{1.8}{$c/a$}}
\psfrag{ca}{\scalebox{2.4}{$c/a$}}
\scalebox{.4}{\includegraphics{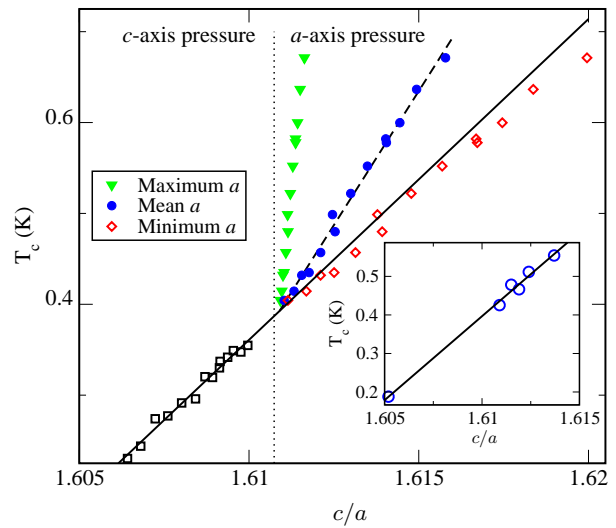}}
\caption{$T_c$ vs. $c/a$. For $a$-axis pressure, we calculate $c/a$ using the
minimum, geometric mean, and maximum $a$ values; see text for details.  The
solid line is a least-squares fit to the $c$-axis data, with slope 35 K.   The
dashed line is a least-squares fit to the $a$-axis data using the mean values
for $a$.  Inset: putative $T_c$ vs. $c/a$ from substitution and
pressure data, with hybridization potential $V_{pf}$ equal for all points.}
\label{tcvsca}
\end{center}
\end{figure}
 
Figure \ref{tcvsca} redisplays the data of Figure \ref{tcvsp}
to $T_c$ as a function of $c/a$.  We use room temperature lattice
constants \cite{yoshinori} and the elastic constants of CeRhIn$_5$
\cite{rkumar692004} to compute $c/a$ at different pressures
\cite{yoshinori}.  We assume that the CeIrIn$_5$ elastic constants are
similar \cite{oeschler}.  Also, at our low to moderate pressures, the
stress/strain relations are still linear. 

\begin{table*}[bt]
\caption{Values used in adjusting for effect of hybridization changes on $T_c.$
Zero-pressure structural parameters from \cite{yoshinori, maehira}; hydrostatic
pressure data from \cite{rkumar702004}.  See discussion in text.}
\begin{tabular}{lcccccc}
\hline
\hline\\
&  &  &  & & $T_c$ (K) at&
$c/a$ at\\
& $T_c$ (K) & $c/a$ & $V_{pf}$ & $\Delta V_{pf} (\%)$ & $\Delta V_{pf}=0.2\%$ & 
$\Delta V_{pf}=0.2\%$ \\
\hline
CeIrIn$_5$ (P=0) & 0.40 & 1.6109 & 2.0240 & -- & -- & -- \\
CeIrIn$_5$ (29 kbar) & 1.05 & 1.6099 & 2.1277& 5.1 & 0.4254 & 1.6109\\
CeCoIn$_5$ (P=0) & 2.30 & 1.6368 & 2.0930 & 3.4 & 0.5116 & 1.6124\\
CeCoIn$_5$ (14 kbar) & 2.60 & 1.6435 & 2.1578 & 6.6 & 0.4666 & 1.6119\\
CeRhIn$_5$ (24 kbar) & 2.50 & 1.6270 & 2.1326 & 5.4 & 0.4783 & 1.6115\\
CeIrIn$_5$ (5.17 kbar, $a$-axis) & 0.67 & 1.6158 & 2.0312& 0.35 &0.5543& 1.6137\\
CeIrIn$_5$ (2.42 kbar, $c$-axis) & 0.23 & 1.6063 & 2.0273 & 0.16 & 0.1880 & 1.6052\\
\hline
\end{tabular}
\label{scalenums}
\end{table*}
                                                                                
Just as for measurements across the CeMIn$_5$ family, we find that $T_c$
increases with increasing $c/a$.  The dotted line of Figure \ref{tcvsca}
is at the ambient-pressure value for $c/a$.  Applying $c$-axis pressure
decreases $c$ while increasing $a$ through the Poisson ratio, so the data
from $c$-axis pressures appear to the left of the dotted line. On the
other hand, $a$-axis pressure decreases the $a$ lattice constant along
the pressure direction and increases $c$, but also increases the $a$-axis
lattice constant perpendicular to the pressure direction. Since the
largest change in lattice constant is along the pressure direction, the
overall effect is an increase in $c/a$, and the $a$-axis data appear to
the right of the dotted line.  However, for $a$-axis pressure the ratio
$c/a$ is not a single well-defined quantity. The value of $a$ is minimum
in the direction parallel to the applied pressure and maximum
perpendicular to the pressure.  Both of these extreme values are used to
calculate $c/a$, with the results plotted in Figure \ref{tcvsca}.  The
remaining data set in Figure \ref{tcvsca} uses the geometric mean of the
maximum and minimum $a$ values, which is the natural way to compare
in-plane areas with the perpendicular lattice constant.  As seen in
Figure \ref{tcvsca}, using the mean value gives a kink in $dT_c/d(c/a)$
at the crossover from $c$-axis to $a$-axis pressure. The larger slope of
the $a$-axis data may indicate the influence of hybridization. Since the
hybridization depends on the spacing between atoms, it increases for both
directions of uniaxial pressure.  However, the increase is small because
the decrease in atomic spacing along the direction of the applied
pressure is partly compensated by increased spacings in the perpendicular
directions.

By considering several different methods of shifting $T_c$, we can
separate the true influence of geometry on the transition temperature
from the effect of hybridization.  To do this, in Table \ref{scalenums} we
consider the effects of hydrostatic pressure, chemical substitution, and our
$a$-axis and $c$-axis pressure measurements.  We use Harrison's
calculation \cite{harrison83} of $V_{pf}$ in a tight
binding approximation,
$\eta_{pf}\frac{\hbar^2}{m}\frac{\sqrt{r_pr_f^5}}{d^5}$. Here
$\eta_{pf}=10\sqrt{21}/\pi$ is a dimensionless constant, $m$ is the electron
mass, and $r_f=0.445$\AA\ and $r_p=19.1$\AA\ are wave function radii
\cite{rkumar702004, harrison}. For $r_p$ we extrapolate values from
\cite{harrison}, p. 644.  For hydrostatic pressure data, we take the
percentage changes in $V_{pf}$ and $c/a$ from Table II and Figure 2  of
\cite{rkumar702004}.

Each experiment in Table \ref{scalenums}---substitution, applied
pressure, or a combination of the two---increases $V_{pf}$ over its value
in CeIrIn$_5$ at ambient pressure, with a percent change given by $\Delta
V_{pf}$. For each type of measurement we simultaneously scale the changes
in $T_c$, $c/a$, and $V_{pf}$ by a single factor to reach $\Delta
V_{pf}=0.2\%$. Our scaled values show how $T_c$ would vary as a function of
$c/a$ with $V_{pf}$ held constant.  The results from the final two columns of
Table \ref{scalenums} also appear in the inset of Figure \ref{tcvsca}. The
solid line is a best fit to all six points, with slope 44 K.  Significantly,
the slope remains 44 K if the fit omits the $c$-axis pressure data, which has
particularly small hybridization change and lies far from the other points.

The broken symmetry perpendicular to the $c$-axis when pressure is
applied along the $a$ axis has little effect on the
superconductivity, judging from the linearity of the fit in the Figure
\ref{tcvsca} inset. We note that using the minimum value of the lattice
constant $a$ preserves the linearity of $T_c$ vs. $c/a$.  Although
probably coincidental, this may indicate how the destruction of
tetragonal symmetry affects the superconductivity.

Our measurements show that up to a few kbar $T_c$ changes linearly with
uniaxial pressure, at 56 mK/kbar for $a$-axis pressure and -66 mK/kbar
for $c$-axis pressure.  This confirms the important role of $c/a$ in
controlling the onset of superconductivity.  Changes in $fp$
hybridization between the Ce and neighboring In atoms also have a large
effect. By comparing several techniques that alter the dimensionality and
the hybridization to different degrees, we control for hybridization
changes and find a pure geometric influence of $\partial
T_c/\partial(c/a)=44$ K.

We thank S. Johnson for useful discussions and J. Ma for helping to
polish the samples.  We acknowledge support from NSF under DMR-0454869 and
DMR-0802478.

*Present address: Quantum Design, 6325 Lusk Boulevard, San Diego, CA.

\end{document}